\documentclass[12pt,preprint]{aastex}

\usepackage{graphicx,amsmath,natbib}

\slugcomment{Accepted to ApJ}

\shorttitle{Temporal evolution of coronagraphic dynamic range}
\shortauthors{Hinkley et al.}

\begin{document}

\title{Temporal Evolution of Coronagraphic Dynamic Range, and Constraints on Companions to Vega}

\author{Sasha Hinkley}
\affil{\it Department of Astronomy, Columbia University, 550 West 120th Street, New York, NY  10027}
\email{shinkley@astro.columbia.edu}

\author{Ben R. Oppenheimer, R\'emi Soummer\altaffilmark{1,2}, Anand 
Sivaramakrishnan\altaffilmark{1,5}}
\affil{\it Astrophysics Department, American Museum of Natural History\\
                Central Park West at 79th Street, New York, NY 10024}

\author{Lewis C Roberts Jr.}
\affil{\it The Boeing Company, 535 Lipoa Parkway, Suite 200, Kihei, HI 96753}

\author{Jeffrey Kuhn}
\affil{\it Institute for Astronomy, University of Hawaii, 2680 Woodlawn Drive, Honolulu, Hawaii 96822}

\author{Russell B. Makidon\altaffilmark{1}}
\affil{\it Space Telescope Science Institute, 3700 San Martin Drive, Baltimore, MD 21218}

\author{Marshall D. Perrin\altaffilmark{3}}
\affil{\it Department of Astronomy, 601 Campbell Hall, University of California Berkeley, CA 94720}

\author{James P. Lloyd}
\affil{\it Astronomy Department, 230 Space Sciences Building, Cornell University, Ithaca, NY 14853}

\author{Kaitlin Kratter\altaffilmark{4}, Douglas Brenner}
\affil{\it Astrophysics Department, American Museum of Natural History\\
                Central Park West at 79th Street, New York, NY 10024}



\altaffiltext{1}{NSF Center for Adaptive Optics.}
\altaffiltext{2}{Michelson Post-doctoral Fellow.}
\altaffiltext{3}{Michelson Graduate Student Fellow.}
\altaffiltext{4}{Current address: Department of Astronomy, University of Toronto, 60 St. George Street, Toronto, ON M5S 3H8}
\altaffiltext{5}{Stony Brook University}


\begin{abstract}
The major obstacle to the direct detection of companions to nearby stars 
is the overwhelming brightness of the host star. Current instruments 
employing the combination of adaptive optics (AO) and coronagraphy can 
typically detect objects within 2$^{\prime \prime}$ of the star that are $\sim 10^{4-5}$ times fainter.
Correlated speckle noise is one of the biggest obstacles limiting such 
high-contrast imaging. We have obtained a series of 284 8 s,  
AO-corrected, coronagraphically occulted $H$-band images of the star Vega 
at the 3.63 m AEOS telescope located on Haleakala, Hawaii. 
This dataset is unique for studying the temporal behavior of speckle noise, 
and represents the first time such a study on highly corrected {\it coronagraphic} AO 
images has been carried out in a quantitative way. 
We find the speckle pattern to be highly stable in both position and 
time in our data. This is due to the fact that the AO system corrects 
disturbances to the stellar wave front at the level where the instrumental 
wave front errors dominate. Because of this, we find that our detection 
limit is not significantly improved simply with increased exposure time alone. 
However, we are able to improve our dynamic range by 1.5-2 magnitudes 
through subtraction of static/quasi-static speckles in two rotating frames: the telescope pupil
frame and the deformable mirror frame. 
The highly stable nature of speckles will exist for any program using a combination 
of coronagraphy and high-order AO, and underscores the importance of 
calibration of non-common path errors between the wave front sensor and the 
image plane. Such calibration is critical for high-contrast AO systems and we 
demonstrate this using empirical data. From our data, we are able 
to constrain the mass of any purported companion to Vega to be less than $\sim 45 M_J $
at $8$ AU and less than $\sim 30 M_J$ at 16 AU, radii not previously probed at these sensitivities. 
\end{abstract}




\section{Introduction}
While the indirect detection of more than 200 planets over the past 
decade \citep{mpv05,mbf05} has been a significant accomplishment for 
exoplanetary science, a current major thrust for the field is the direct 
detection of these objects through high-contrast imaging. Very large 
telescope apertures, although extremely effective for gathering photons 
\citep{l02}, are not a necessity for direct imaging \citep{osm03}. The key requirement is 
the suppression of the host star's overwhelming flux. A ``hot jupiter'' 
will typically be $10^7$-$10^9$ times fainter than its host 
\citep{bcb03,b05} and often situated within a fraction of an 
arcsecond of the star. A promising method for direct imaging 
of stellar companions involves two techniques working in conjunction: 
high-order adaptive optics (AO), providing control and manipulation of the 
starlight \citep{a94}, and an optimized Lyot coronagraph \citep{skm01,odn04} to 
suppress this light. Together, these two techniques can obtain 
contrast levels of $10^5$ or better within $2^{\prime\prime}$. After this level of contrast has 
been achieved, the major source of noise in the images is correlated 
speckle noise. Without a coronagraph, \citet{rwn99} has demonstrated that 
speckle noise will dominate over photon noise by a factor $\sim10^4$. 
With a coronagraph, the contribution from the ``pinned'' speckle noise can 
be significantly reduced \citep{slh02,psm03,as04}.  
Such speckle noise is largely due to non-common path 
errors (those not measured by the wave front sensor) e.g. small 
aberrations in the coronagraphic optics. Other sources
include AO correction errors such as fitting error, aliasing, lag \citep{jvc06} and Fresnel wavefront propagation effects. 

The Lyot Project (Oppenheimer et al.\ 2003; Oppenheimer et al.\ 
2004; Oppenheimer et al.\ 2006a, in preparation) employs an optimized, 
diffraction-limited Lyot coronagraph \citep{l39,m96,skm01}. It is deployed at the 
$D=3.63$ m Advanced Electro-Optical System (AEOS) telescope in Maui, 
with an AO system equipped with a 941 actuator deformable mirror (DM). 
The telescope is an altitude-azimuth design, with a beam traveling under the dome 
floor to a coud\'e room containing the AO system and the DM  \citep{rn02}. During an observation, the DM frame of reference is always aligned with the frame of our infrared camera. The frame of the 
telescope pupil containing the secondary mirror support struts (the ``spiders''), however, rotates with respect to the camera during an observation. For the purposes of this study, we mention that these frames contain two identifiable sources of speckle noise in our observations.

In March 2004, the project began a survey of approximately 100 nearby stars in 
coronagraphic mode in the $J$,$H$, and $K_s$ bands with the goal of 
detecting faint companions and disks orbiting the stars. On 2005 May 14 (UTC), we 
obtained a sequence of 284 AO-corrected, coronagraphically occulted 
images of the bright star Vega (HD172167, A0V, $V=0^m$). 
We estimate our uncorrected wavefront errors at $\sim 150$nm, and we observed with adequate 
seeing ($r_0=14.1 \pm 2.5$cm).
With {\it atmospheric} phase errors well under control by the AO, any 
speckle noise should come largely from the AO system or the coronagraph 
itself. Such noise largely arises from non-common path errors between the wave front sensor 
and the image plane, and can be mitigated through careful calibration 
\citep{wgs04, hbf03,bfh03}.  Although quasi-static speckle noise has been shown 
to be the main limitation to earlier instruments \citep{mdn05,bcl03,brb04} this dataset is 
unique for studying the temporal characteristics of correlated speckle noise in such a 
highly corrected system coupled to an optimized Lyot coronagraph.

After describing our observations (\S2), we will describe our method 
for quantifying our sensitivity in the following section. We will show that our sensitivity does 
not increase substantially when many images are simply coadded together, 
but demonstrate  that the dynamic range can be improved by subtracting those speckles
that are static in the telescope pupil frame and also those that are fixed with respect to the DM. 

\section{Observations}
We obtained a sequence of 284 AO-corrected, 8 s images of the star Vega in the 
$H$-band (1.65 $\mu$m) with the AEOS 3.63 m telescope on 2005 May 14 (UTC). 
The AO system's 941 actuator deformable mirror is complemented by a tip/tilt loop 
capable of running up to $\sim 4$ kHz, a Shack-Hartmann wave front sensor with 
a 2.5 kHz frame rate, and a real-time 
wave front reconstructor using least-squares calculations performed on 
dedicated hardware \citep{rn02}. This combination of features has the 
potential for some of the highest-order correction in modern AO. For all 
observations, the AO loop was closed and all images were fully occulted 
using our focal plane mask with a 455 mas diameter 
$(4.9\lambda/D)$. Although the theoretical $H$-band full-width at half 
maximum (FWHM) is 94 mas for AEOS ($D=3.63$ m), the effective FWHM is 
121 mas when the size of the Lyot stop is taken into account ($D_{Lyot} 
= 2.81$ m). The images were recorded with the Kermit IR Camera 
\citep{p02} in the $H$-band with a 13.5 mas/pixel 
platescale. During the observations, the star covered an elevation 
range between $\sim72^{\circ}$ and $\sim67^{\circ}$, and thus suffered a 
differential atmospheric refraction offset \citep{r02} of 60 mas 
between its $H$-band and its $V$-band (0.55 $\mu$m) position 
on the sky.
To correct for this, we applied offsets ($\sim 10$ mas each) to the
tip-tilt mirror to obtain the centered $H$-band occulting position. 
In addition, our Lyot coronagraph has the capacity for 
simultaneous dual beam polarimetery, with the goal of obtaining Stokes 
$I$, $Q$, and $U$ images (Oppenheimer et al.\ 2006b, in preparation). All 
of the data analysis for this paper was performed on Stokes $I$ images (the 
measure of total intensity). We will save the in-depth analysis of the polarimetric sensitivity 
for another work. 

\section{Analysis \& Interpretation}
\begin{figure}[ht]
\center
\resizebox{1.0\hsize}{!}{\includegraphics{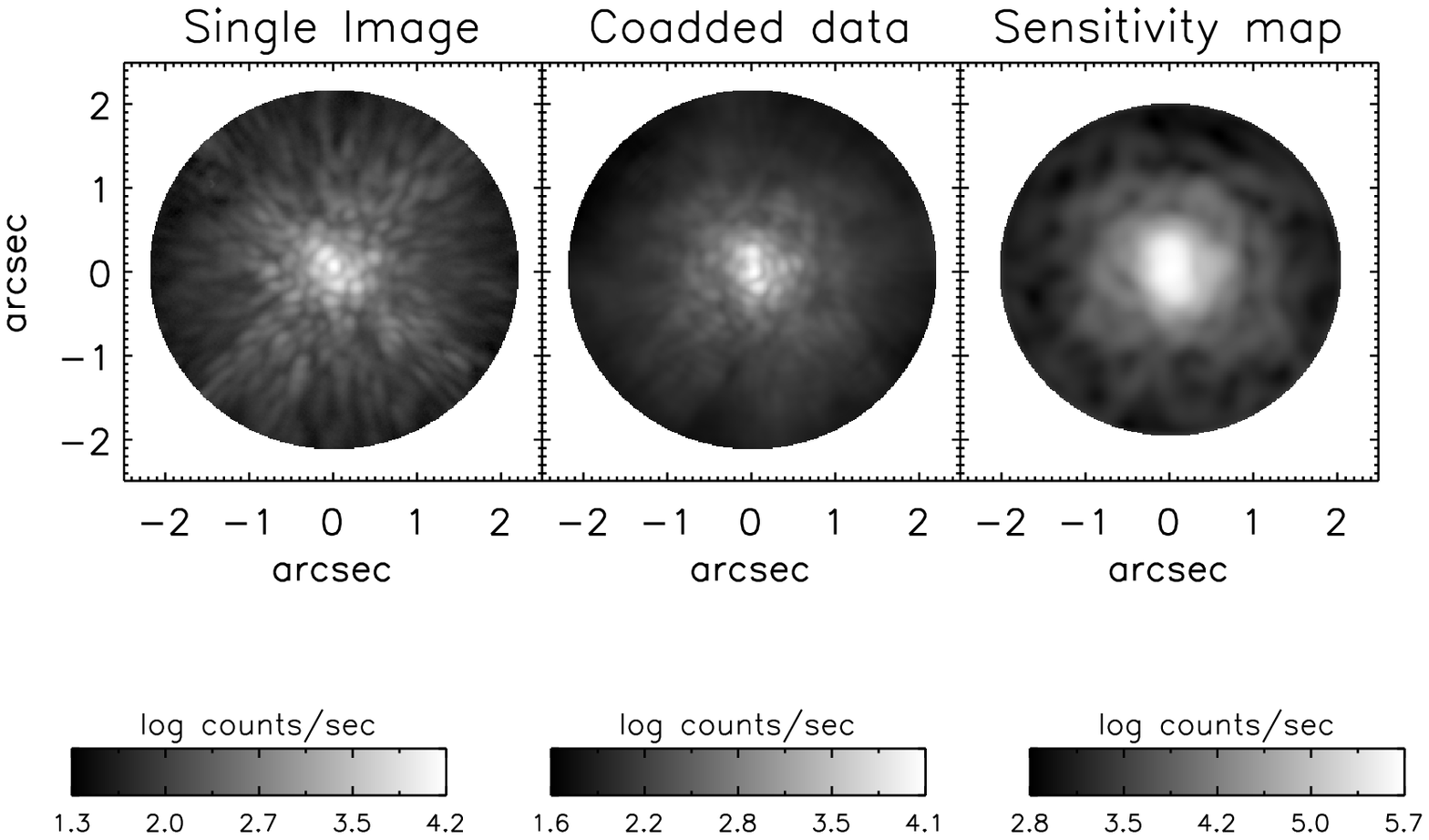}}
  \caption{Left: This image shows a single $H$-band 8 s exposure of the star Vega. The occulting coronagraphic mask is in place and the AO control loop is fully closed. The speckle pattern which limits the dynamic range is evident, and note that residual pinning is present at regions close to the center, as well as on the telescope ``spiders''. Also, the first Airy rings are not completely removed by the Lyot coronagraph. Middle: This image shows all 284 exposures coadded together.  Right: the corresponding 5$\sigma$ detection limit map constructed from the middle image. Each pixel in this figure represents the minimum flux necessary for a $5\sigma$ detection of a companion. For each image, the platescale is 13.5 milliarcseconds/pixel.}
  \label{sigma}
\end{figure}

The raw data images required a mix of both traditional data reduction steps as well as some techniques customized for the Kermit Camera. Each of the 284 raw images were cropped to a region containing the star, dark subtracted, flat field corrected, and cleaned for bad pixels and cosmic rays through interpolation. Due to cross-talk within the detector electronics, the raw data also contained easily-removable negative electronic ``ghost'' images of the occulted star about 16 times less intense at 128-pixel intervals from the true position of the star.
Each image was then rotated so that the image $y$-direction is coincident with North on the sky. After the basic reduction steps were complete, 284 new coadded images were formed in which the $n^{th}$ coadded image was the mean of the first $n$ individual images. 
The image representing the mean of 284 images is shown in the middle of Figure~\ref{sigma}. 


The dynamic range or contrast at a given position in a two-dimensional 
image is commonly quantified as the faintest companion detectable at that 
location at the 5$\sigma$ level \citep{osm03}. We use this convention. 
In these data, the main noise contribution is due to the speckles, and 
has to be estimated locally, from the data itself. 
In addition, given that a typical PSF in these data has an 
oversampling factor of 7-9 within $\lambda/D$, the signal should not be 
measured from a single point (e.g.\ the brightest pixel) but rather should 
take into account {\it all} the pixels within one resolution element. So, at a given 
location, calculating the dynamic range consists of measuring the signal 
from a hypothetical companion by integrating its flux over a patch of size 
$\lambda/D$. To remain consistent, this signal needs to be compared to the 
noise in the same integrated area if no real PSF exists at that location, 
or in the immediate adjacent area of the image. 
There are alternatives to this particular method of calculation, but 
they agree to within 0.3 magnitudes \citep{soh06}. 

We constructed a detectability map with the same number of pixels and 
pixel scale as each science image, indicating the image's $5\sigma$ detection limit as a 
function of position in the image. The corresponding $5\sigma$ map for 
the $n=284$ image is shown on the right of Figure~\ref{sigma}. Each 
pixel in such an image represents five times the root-mean-square 
variation of the pixel values 
in a circular subregion centered on the corresponding pixel in the 
occulted image. The subregion was chosen to have a $0.28''$ 
diameter (somewhat larger than the $H$-band point spread funtion). 
Each pixel in such a detection limit map represents the minimum flux 
necessary for a $5\sigma$ detection in the co-added image. Thus, these 
maps show the dynamic range of the image. 


Due to Vega's brightness, unocculted calibration images of it were not available to determine apparent magnitudes. Thus, each of the detection limit maps was calibrated to a fainter, 
unocculted standard star (HD160346, K3V, $V=6.5^m$) observed at nearly the same time and airmass as Vega. 
The photometric zeropoint was determined by 
comparing the calibration star's tabulated $H$-band magnitudes with its flux 
counts. With this value in hand, image counts could then be converted 
to apparent magnitudes. We calculated an $H$-band zeropoint of 20.18 
magnitudes and use this value for all further analysis in 
this paper. In addition, a short sequence of images of this same star was used to derive $H$-band Strehl ratios near 60\%. However, during the Vega observations, the AO system was likely performing somewhat better than this, given Vega's brightness. 

In order to determine our detection limits, we 
studied how the dynamic range and structure of this map changes with the 
value of $n$, the number of images added together. This is equivalent to 
studying dynamic range as a function of total exposure time. To do this, we 
calculate azimuthally-averaged dynamic range values at many different 
radial positions in the image. 
We plot these values (dashed  lines) as a function of exposure time for two 
different radii in Figure~\ref{vsimages}. 

\begin{figure}[ht]
\center
\resizebox{1.0\hsize}{!}{\includegraphics{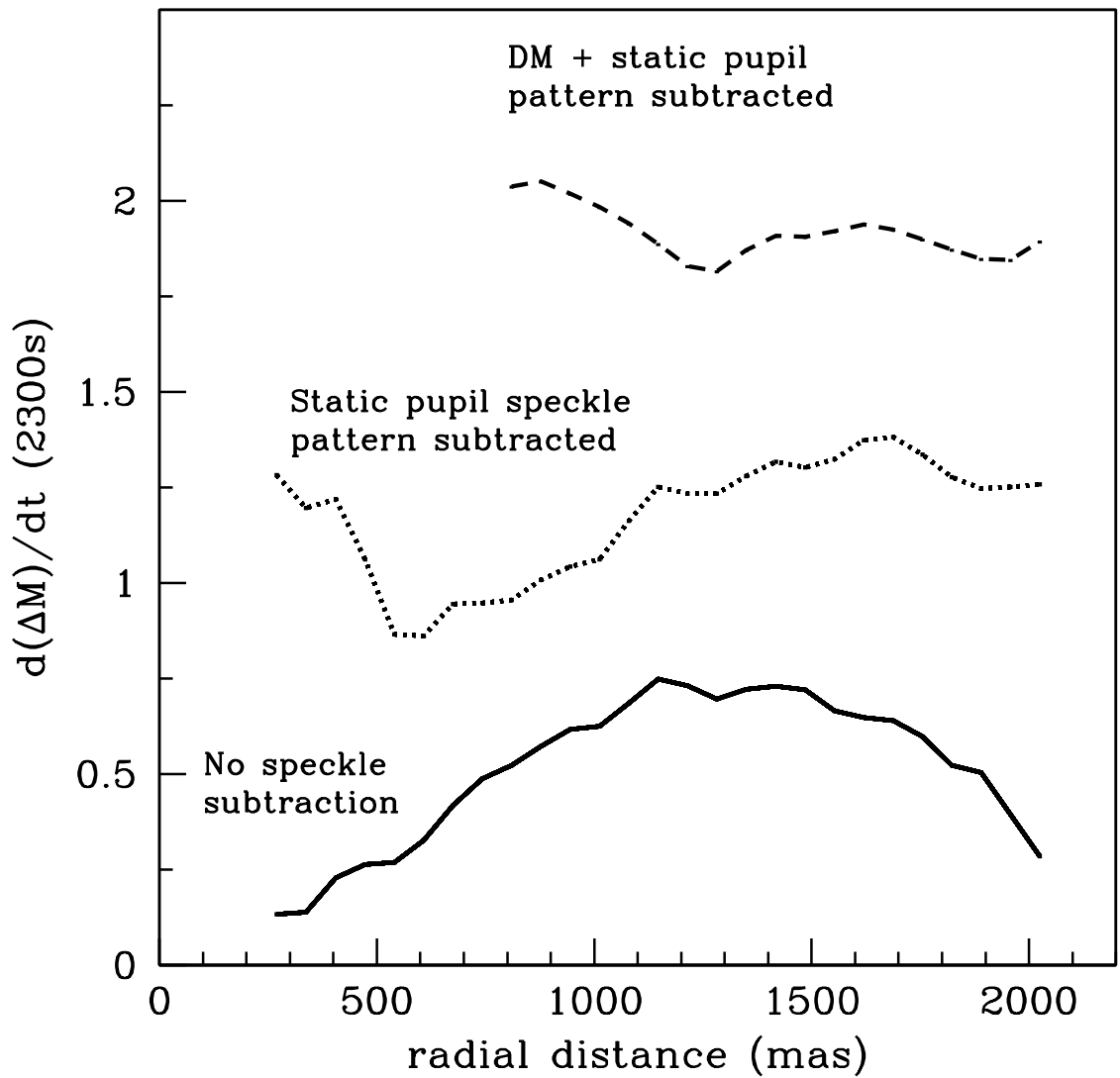}}
\caption{This plot shows the gain (minimum to maximum) in dynamic range over the full $\sim2300$s observing sequence as a function of radial distance from the star. The solid line shows the increase by simply coadding the images together, and only marginal improvement is obtained due to the highly static nature of the speckle pattern. When the truly static contribution in the frame of the telescope pupil is subtracted out (dotted line), the dynamic range gain improves. Finally, when the speckle pattern due to the imperfections in the DM are also removed (dashed line), an increase of $\sim2$ magnitudes can be acheived over the course of the entire observing sequence. This last dynamic range increase is only valid beyond $\sim700$ mas, due to the inner working angle defined  by the rotation in this frame.  }
\label{drslope}
\end{figure}

It is immediately apparent that these curves are flat. They do not 
seem to reach substantially fainter magnitudes with longer effective integrations. 
Any further gain in sensitivity through subsequent observations seems 
to be essentially negligible. The gain in dynamic range over time does depend on the radial position in the image as shown in Figure~\ref{drslope}. 
At 300 mas, the increase over the course of 2300 s is only $\sim 0.2$ magnitudes. 
However, at 1500 mas, the increase over the same time period increases to $\sim 0.7$ magnitudes.)  
A certain degree of the increased dynamic range within this region may be due to the AO control 
cutoff, given by $\lambda N_{act}/2D $ (Sivaramakrishnan et al. 2001; Oppenheimer 
et al. 2003, 2006a, in preparation; Poyneer \& Macintosh 2004), where $N_{act}$ 
is the number of linear actuators across the pupil (34 for AEOS). For AEOS, this control cutoff is $\simeq1500$ mas, near where the solid curve in Figure~\ref{drslope} peaks. 

This relative plateau in sensitivity is due to the 
inherently correlated nature of the speckle structure from frame to frame, 
leading to high spatial and temporal stability. Since the amplitude of 
this variation in intensity is highly stable across an image \citep{mdn03}, 
a number of short exposures added together will not cause the noise to 
average out as e.g.\ \citet{rwn99} has suggested for the case of atmospheric speckles. 
Without subtracting out this source of noise, Figure~\ref{vsimages} shows that the fixed speckles 
dominate the overall dynamic range.

\subsection{Subtraction of static speckles}
The static and quasi-static speckles that limit the dynamic range arise primarily from two sources: those speckles pinned to the diffraction pattern of the secondary support struts (the ``spiders'') in the telescope pupil, and those that are due to the quasi-static imperfections in the AO optics. The pupil frame will rotate with respect to the infrared detector, while the AO frame (DM) is fixed relative to it during an observation. To get a handle on the degree of speckle stability, we performed two experiments on the data, each an attempt to correct for the two sources. We performed the experiments on the data after its pre-processing, but prior to forming the coadded images and carrying out the rest of the dynamic range calculation. We address the first of these experiments in this section. 

In order to quantify the degree to which we can subtract out the longest-lived speckles due to the spiders, we ``de-rotated'' each of the 284 processed images by their parallactic angle offset so that the pupil frame pattern had the same orientation in each image. The sky thus rotates in each image. These rotated images were median combined into a single image, and this median image was subtracted from each of its constituent images. This step is identical to the first of two steps carried out by \citet{mld06} to achieve a gain in dynamic range in high-contrast data. Next, all these constituent images were rotated so that North/East were aligned with up/left. With this new data set, the exact same dynamic range calculation was performed (sequential coadds formed, $5\sigma$ detection map formed, etc.) and the results are shown in Figure~\ref{vsimages} (dashed-dot curve). An increase of 0.5-1.0 magnitudes is seen over the case with no speckle subtraction. Moreover, Figure~\ref{drslope} shows that the gain in dynamic range during the nearly 2300 s observation varies from $0.8$ magnitudes nearer to the star ($\sim 500$ mas) to $\sim 1.3$ magnitudes at greater separations ($\sim 1500$ mas). 

It should be noted, of course, that the measured improvement in dynamic range is not valid at all radii. The boost in dynamic range is only valid in those regions beyond which a potential companion in the now rotating sky frame will rotate enough through the sequence so that it does not become part of the median image and is subsequently subtracted out \citep{mld06}. Given the $40^\circ$ rotation of the parallactic angle and assuming a minimum necessary companion separation of 1.5 FWHM over the sequence, this defines an inner working angle of 250 mas, close to the edge of the occulting mask used.  

\subsection{Lifetimes of quasi-static speckles}
Once these truly static speckles were subtracted out of the image, we studied the lifetime of those remaining quasi-static speckles. To do this, we performed a pixel-wise temporal autocorrelation analysis similar to that in \citet{fg06}. 
We calculated the temporal autocorrelation of each pixel over the entire image. The autocorrelation function, shown in Figure~\ref{autocor} for three sample locations in the image,  reveal two distinct time scales. As a convention, we define these time scales by the half-width of the two regions separated by the ``knee'' in the autocorrelation function near 20 s shown schematically in Figure~\ref{autocor}. The first of these, which we denote by $\tau_{short}$, represents any rapid decorrelation of the speckles that may exist. This decorrelation {\it does} correspond to a slight increase in the dynamic range, and this can be seen at the very beginning of the curves in Figure~\ref{vsimages}. The longer lifetime, denoted in the image by $\tau_{long}$, gives a reasonable measure of the lifetime of the quasi-static contribution. A map of both these lifetimes is shown in Figure~\ref{longlifetime}. The maps reveal the localization of the slowest and fastest speckles: the longest lifetimes of quasi-static speckles are $\sim 400$ s and the shortest lifetimes of the short-term speckles are 1 - 5 s.  Although significant work has been done on the lifetimes of {\it atmospheric} speckle lifetimes \citep{rgl82,mps05}, very little formalism exists for static and quasi-static speckle lifetimes.  

Also in Figure~\ref{longlifetime}, we show radial profiles of the speckle lifetime maps. Although no particularly strong radial features are evident in the long-term lifetime map, there still is a notable  increase in the lifetimes near 500 mas, reflecting those relatively static speckles that are pinned just outside the coronagraphic mask. Also, the radial curves reveal why we achieve slightly more dynamic range towards the outer parts of the images. Specifically, the average quasi-static speckle lifetime near 740 mas is $\sim210$ s, while at 1760 mas it is $\sim 150$ s. Averaging the shorter lifetime speckles will achieve more dynamic range increase than averaging longer ones over the same period of time. 
This is well reflected in the curves of Figures~\ref{vsimages} and \ref{drslope}, where the increase in dynamic range with integration time is $\sim 0.8$ magnitudes at 740 mas and $\sim 1.4$ magnitudes at 1760 mas.  

In the short-term plot, the lifetimes of those speckles closer to the star are smaller. Although a definitive physical mechanism underlying this remains unclear, this behavior is reminiscent of the discussion in \citet{mps05}, in which simulations of pinned speckles---in our case, residual pinned speckles close to the star---fluctuate quickly with a period dictated by the wind speed carrying the wave front error. The type of approach shown in Figure~\ref{longlifetime} may serve as a useful tool for future work aiming to further explore the physical nature behind speckle lifetimes. 

\subsection{Subtraction of DM speckles }
After the subtraction of the truly static speckles in the frame of the telescope pupil, an attempt was made to subtract out another contributor to the speckle pattern: those speckles caused by static aberrations in the adaptive optics system, coronagraph, or science camera. In a similar manner to that described in Section 3.1, the pupil frame subtracted images (north aligned up, east aligned left) were de-rotated to the frame in which the placement of the DM is fixed with respect to the IR array. This is  the same frame of reference occupied by the images when they initially come straight from the infrared camera. Similarly, these images were median combined into a single image and this image was subtracted from each of the images in the DM frame. The newly subtracted images were rotated back to their normal ``sky'' coordinates (north-up and east-left) and we calculated the dynamic range in exactly the same way. 

The evolution of the dynamic range after both (pupil frame and DM frame) subtractions is shown in Figure~\ref{vsimages} (solid line). Depending on the location in the image plane, the gain achieved with these two subtractions over simply coadding images (no subtraction at all) is $\sim1.5$-2.0 magnitudes. The increase in sensitivity over different radial positions in the image is shown more fully in Figure~\ref{drslope}. For this second case, we are not allowed as generous an inner working angle. The transformation between the ``sky'' frame and the ``DM'' frame only spans $12^\circ$ over the  sequence of 284 images. This rather small angle corresponds to an angular separation of $\sim700$ mas outside of which the PSF of a true companion will not be incorporated into the median image and subsequently subtracted. Hence, improvements to the dynamic range due to this method are only valid outside this separation. Because of this, an observing sequence which incorporates the maximum amount of field rotation will benefit this kind of data reduction.

\subsection{Dynamic Range}
Figure~\ref{vsradius} shows azimuthally-averaged radial plots of our
dynamic range incorporating all images for each of the three cases: no speckle 
subtraction, pupil frame speckle subtraction, and both pupil+DM speckles subtracted. 
It should be noted that for the case of no speckle subtraction, the dynamic range will still 
contain some of the highly non-axisymmetric structure due to the telescope ``spiders''. 
However, the curves for the two speckle-subtracted cases do not suffer as much from this, and the 
azimuthally averaged curves are more justified. 
The speckle noise will be most prominent close to the star where 
residual speckle pinning persists due to the insufficient classical Lyot coronagraph starlight suppression 
\citep{bdt01,slh02,as04}. An upgrade using an apodized pupil Lyot coronagraph \citep{s05} will  help the dynamic range in this region. At $\sim 2$ arcseconds, our $H$-band detection 
limit is approaching $\Delta M \gtrsim$ 12.5, where the speckle noise has 
diminished in strength. 

The figure also shows the sensitivities of VLT/NACO observations incorporating the SDI technique \citep{bcm06,kab05}, as well as the Gemini Altair observations with the Angular Differential Imaging technique \citep{mld06} applied to them. Although the other projects' curves are for stars other than Vega, and detection diagnostics in each of these systems will be different and subject to peculiarities associated with each (non-uniformity in exposure times, different wavelengths and telescopes, etc.), these curves still place a valuable benchmark for our sensitivity. We feel that these two projects are strong representations of the current state-of-the-art in high dynamic range imaging. Our results demonstrate that a 3.63 m telescope with
a high-order adaptive optics system and a coronagraph can provide comparable contrast to that achievable with 8 m telescopes. This is highly promising for future high-contrast AO systems on 8-10 m telescopes.  

However, the relative lack of substantial ``halo-clearing'' within the AO control cutoff also arises from several limiting factors intrinsic to the AO system itself. For one, the AEOS deformable mirror 
(DM) was hampered by five malfunctioning actuators. \citet{o05} have shown 
that just a single malfunctioning actuator in a DM can effectively contribute 
enough aberration to significantly increase the quasi-static speckle noise, 
thereby limiting high-contrast work. Due to the 
presence of these broken actuators, the DM was operated with only half of 
the actuator stroke it was designed for, to ensure that no additional 
actuators are damaged. Operating in this mode prevents the 
system from correcting the full tilt of the wave front over a given 
subaperture, especially in poor seeing. Moreover, the AEOS wave front 
is not spatially filtered \citep{pm04} before wave front sensing, which may cause its 
high-frequency spatial components to be interpreted as low-frequency 
components by the wave front reconstructor. This results in degraded 
performance. All of these effects can lead to the creation of a 
prominent speckle pattern, which supersedes the ``halo-clearing'' effect 
of the AO control radius. Tests using a Zygo interferometer have shown 
that the wave front error induced by the coronagraph alone amounts to 
no more than $58$ nm RMS \citep{o05}. By itself, this error will only diminish 
the Strehl ratio to 92\%. This further suggests that the highly stable noise is 
arising from other speckles/scattered light from the 
telescope/AO system and not the coronagraphic optics. 

\section{Limits on Vega's companion}
Earlier studies on the inner $\sim20$ arcseconds of the Vega system have revealed asymmetric dust patterns suggesting the system may possess a planet on the order of a few Jupiter masses ($M_J$) with 5-10 arcsecond separation on an eccentric orbit \citep{hgz98,whk02}. Recent high contrast imaging work  by \citet{mld06} and \citet{hhs06} use $M$ and $1.58 \mu$m imaging to place upper limits of 7 $M_J$ beyond 2.5 arcseconds and 3 $M_J$ beyond 8 arcseconds, respectively. This complements earlier work by \citet{mhw03} and \citet{mbk03} who used $H$ and $K$-band imaging to place an upper limit of 20 and 10 $M_J$, respectively, for  a hypothetical companion located near 7 arcseconds.  

\begin{figure}[ht]
\center
\resizebox{1.0\hsize}{!}{\includegraphics{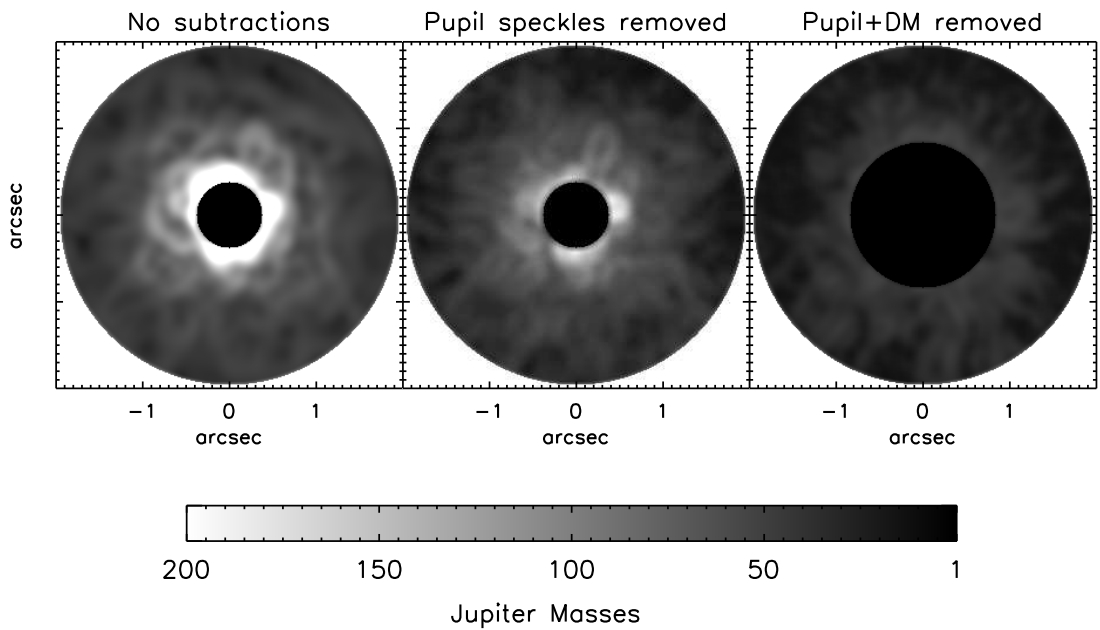}}
\caption{Dynamic range maps incorporating all 284 images.  Each map shows the upper limit on companion masses (in units of Jupiter masses) based on the models of \citet{bcb03}. The left map shows the detectable companion masses with no speckle subtraction whatsoever. The middle plot shows the sensitivity after subtracting the speckle pattern in the telescope pupil frame, while the right plot shows the limits after both the pupil frame pattern and those caused by the imperfections in the DM have been removed.  The larger black circle in the last plot shows the inner working angle ($\sim700$ mas) beyond which the dynamic range measurements are valid when the DM speckles are removed (see text). }
\label{planetmaps}
\end{figure}

\begin{figure}[ht]
\center
\resizebox{1.0\hsize}{!}{\includegraphics{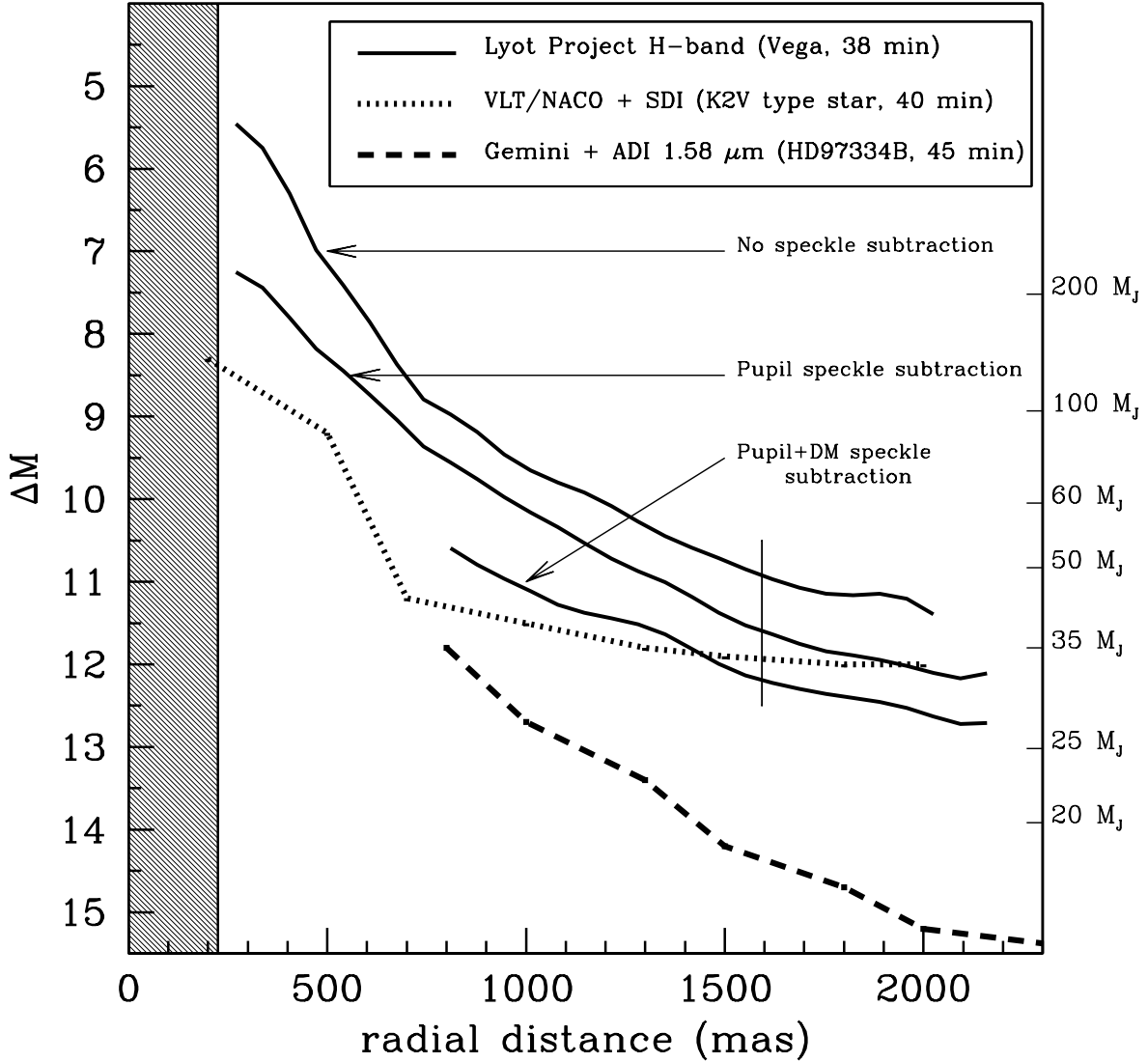}}
\caption{Azimuthally-averaged radial profiles of our $H$-band detection 
limit (solid lines) incorporating all 284 images. The curves show our sensitivity with no speckle subtraction whatsoever, after the static pupil speckle pattern has been subtracted, and after those speckles due to the imperfections in the DM have further been subtracted. 
The shaded region at left represents the radius of our occulting mask. 
The dotted line represents the VLT/NACO sensitivities with the Simultaneous Differential Imaging (SDI) analysis technique \citep{bcm06,kab05}. The dashed line shows the Gemini results from \citet{mld06} using the ADI technique on the star HD97334B. 
The vertical line at 1593 mas shows the $\lambda N_{act}/2D$ extent of the AEOS AO control radius \citep{osm03}. The corresponding upper limits to companion masses at right are based on models from \citet{bcb03} and apply only to the Lyot Project results. The other programs may have different mass limits, given their sensitivities to e.g. methanated companions.
}
\label{vsradius}
\end{figure}

The real value of our current work is our ability to constrain the mass of any unseen massive companions within 2 arcseconds, corresponding to $\sim16$AU for the Vega system. Figures~\ref{vsimages} and \ref{vsradius} show the limiting companion masses at the contrast level achieved. These were calculated by matching our absolute magnitude detection limits with the models of \citet{bcb03} assuming an age of 300 Myr for the Vega system. Since we detect no companions in our images, Figure~\ref{vsradius} indicates we can rule out any companions more massive than about 45 $M_J$ at 1 arcsecond (8AU) and about 30 $M_J$ at 2 arcseconds (16AU). Figure~\ref{planetmaps} shows a two-dimensional representation of the upper limits to detectable companion masses for each of the three methods of coadding: with no speckle subtraction whatsoever, after the pupil-frame speckle subtraction, and after both the pupil-frame and DM speckle subtraction. The increase in sensitivity towards lower masses is evident after the subtraction of the static/quasi-static speckles. Moreover, we compare our mass limits at different angular separations to that for several other recent imaging studies of Vega's purported companion in Table 1.

 \begin{deluxetable}{lllcccccc}
\tabletypesize{\scriptsize}
\tablecaption{Sensitivity limits (in Jupiter masses) at different angular separations for several selected prior imaging studies of Vega. }
\tablewidth{0pt}
\tablehead{\colhead{Work } & \colhead{Wavelength}  & \colhead{Telescope, Technique}     &\colhead{0.5$^{\prime \prime}$ } &   \colhead{1$^{\prime \prime}$ } & \colhead{2$^{\prime \prime}$ } & \colhead{4$^{\prime \prime}$ } & \colhead{7$^{\prime \prime}$ } & \colhead{10$^{\prime \prime}$ }    }
\startdata                                     
                      \citet{mbk03}        &               $K$                    &       Keck, AO                                           &        -          &         -          &          -          &        -              &   10               &     8               \\ 
                     \citet{mhw03}        &               $H$                    &    Palomar, AO                                        &        -          &         -          &          -          &       30           &   15-20            &   $\sim 12$                   \\ 
                      \citet{mld06}         &               $1.58\mu$m     &     Gemini, ADI                                        &        -          &         -          &          -          &         5           &      4                  &     3               \\ 
                     \citet{hhs06}          &               $M$                    &       MMT, AO                                           &        -          &         -          &        26         &         7          &      7                  &     7              \\ 
                     \citet{iof06}            &               $H$                    &   Subaru, AO + coron.                            &        -          &         -          &        120       &         7          &     5-10             &     -                       \\ 
                      This work             &               $H$                    &  AEOS, AO + coron.                                &     135        &       43         &         27         &        -           &        -                 &    -                        \\
\enddata
\end{deluxetable}

\section{Conclusions}
\begin{figure}[ht]
\center
\resizebox{1.0\hsize}{!}{\includegraphics{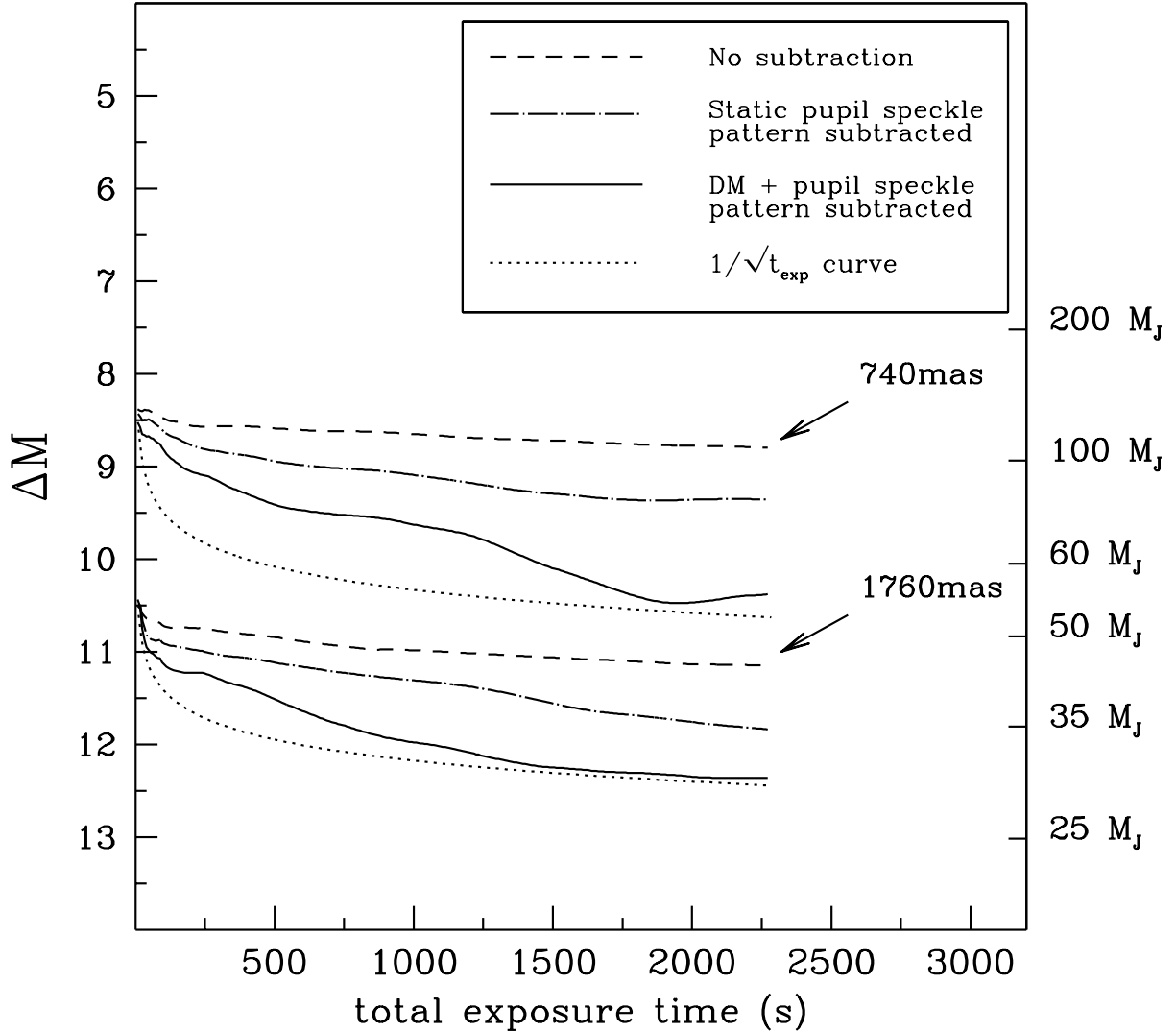}}
\caption{Detection limits as a function of total exposure time for two different radial positions in the field. Each group of lines represents the azimuthally-averaged $5\sigma$ detection limit at the radius listed (in milliarcseconds). 
By simply co-adding images together (dashed line), our dynamic range does not improve substantially with exposure time due to to the presence of highly stable speckles. Subtraction of the truly static portion of the speckle pattern in the telescope pupil frame will improve the dynamic range by nearly a magnitude (dashed-dot line). Finally, another subtraction of those speckles caused by imperfections in the deformable mirror and AO system increases the dynamic range still a bit more (solid line). The dotted line shows the $1/\sqrt t_{exp}$ curve, the expected contrast gain due to uncorrelated noise with a Poisson distribution. }
  \label{vsimages}
\end{figure}

\begin{figure}[ht]
\center
\resizebox{1.0\hsize}{!}{\includegraphics{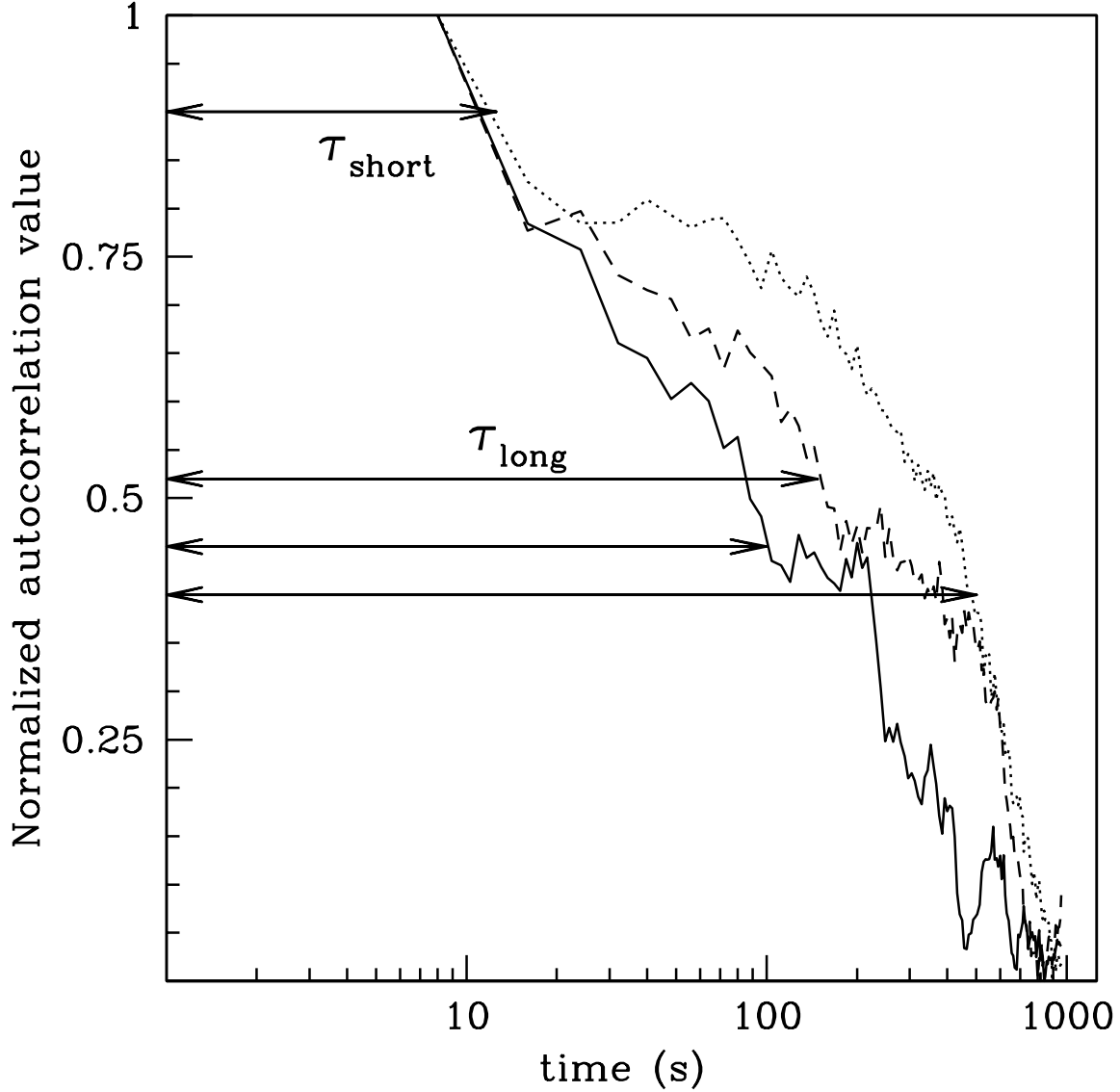}}
\caption{Temporal autocorrelation functions for three different pixel positions located at $0.19^{\prime\prime}$ (solid line), $0.10^{\prime\prime}$ (dashed line), and $.23^{\prime\prime}$ (dotted line) from the star in the image plane. Each of the autocorrelation functions are characterized by two distinct time scales, here defined as the half-width of the two regions separated by the break  near 20s. Each of the three functions have a nearly identical $\tau_{short}$, but three distinct $\tau_{long}$ values. The time axis is shown in a log scale to better illustrate the shorter timescale, $\tau_{short}$. This shorter timescale corresponds to a quick decorrelation of the speckles and the corresponding improvement in the dynamic range can be seen in the earliest parts of the curves in Figure~\ref{vsimages}. The longer timescale, $\tau_{long}$ gives a measure of the lifetime of the quasi-static component to the speckle noise.  }
\label{autocor}
\end{figure}

\begin{figure}[ht]
\center
\resizebox{1.0\hsize}{!}{\includegraphics{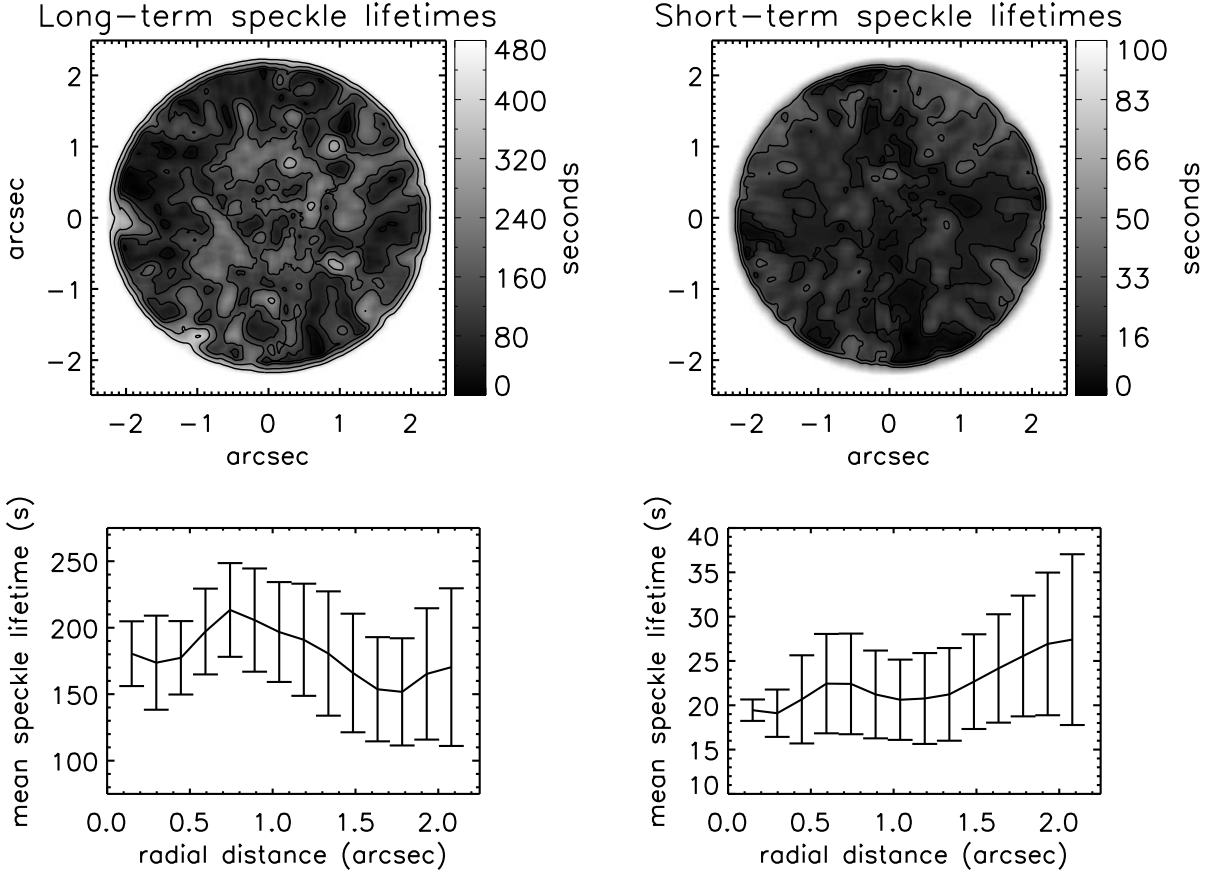}}
\caption{Top panels: maps of the derived lifetimes as shown in Figure~\ref{autocor} for the quasi-static long ($\tau_{long}$) and short-lived ($\tau_{short}$) speckles in the image plane. The lower two panels show azimuthally averaged radial plots of the top panels with $1\sigma$ error bars. The contours correspond to 100, 150, 200, 300, and 400 seconds for the long-term map and 5, 15, 20, 35, and 50 second intervals for the short-term map. The increase in the long-term (quasi-static) lifetimes near 500 mas are due to those speckles that are ``pinned'' just outside the coronagraphic mask. Also, the $40\%$ relative difference in quasi-static speckle lifetimes between 740 mas and 1760 mas is likely responsible for the greater increase in dynamic range at these two locations as shown in Figures~\ref{vsimages} and ~\ref{drslope} during the 2300 s total exposure time. 
}
\label{longlifetime}
\end{figure}

The relative floor in sensitivity shown in Figure~\ref{vsimages} 
is due to the highly correlated and persistent nature of the speckle 
noise in the images. In contrast to the way the random, uncorrelated noise 
is suppressed through averaging, the data are subject to a static and quasi-static 
speckle pattern, placing a limit on the sensitivity of an observing 
sequence. Among other factors, such an influx of speckle noise is likely 
due to a known malfunction in a subset of the AEOS deformable mirror 
actuators. In addition, high-order AO systems in general can suffer from 
a variety of problems (misalignment, mirror figure errors, etc.), each 
of which can be a significant source of speckle noise. We anticipate fixed 
speckles like those discussed here to be a major obstacle for other 
AO+Coronagraphy programs with similar objectives. But more importantly, this 
study underscores the primary importance of non-common path errors between the 
wave front sensor and the science image plane as other authors have discussed \citep{hbf03, bfh03}. Calibration of these errors is 
critical for AO imaging to exceed the $\sim 70\%$ Strehl level. 
We have demonstrated this with empirical data. 

This paper demonstrates the extent to which 
fixed speckles can dominate the overall noise budget of high-contrast AO observations, far exceeding the 
contributions from residual atmospheric effects. 
As we have shown, an increase in sensitivity (1.5-2  magnitudes) can be obtained by 
subtracting off individual contributions to the speckle pattern. Another solution may rely on the wavelength-dependent nature of the speckle noise. 
Although speckles remain fixed in space for a given filter, they will reside 
in different locations in other passbands, especially if they result from 
phase errors in or near pupil planes in the optical train. This key feature 
will allow a faint companion---with a fixed position in all filters---to be 
selected out from the speckle noise. Thus, the addition of multi-wavelength 
observations will enhance the efficiency with which companions may be 
detected \citep{mdr00,sf02,bcl04,mdn05}. Moreover, an integral field unit 
\citep{lqk03} will allow the retrieval of spectra across an image, further 
disentangling companions from the speckle noise. 



\acknowledgments

The authors would like to thank Christian Marois for suggesting the use of the ADI-like multi-plane rotation subtraction technique, Gilles Chabrier and Isabelle Baraffe for providing their models, and Michal Simon for helpful discussions about this paper. 
The Lyot Project is based upon work supported by the National Science  
Foundation under Grant Nos. 0334916, 0215793, and 0520822, as well as  
grant NNG05GJ86G from the National Aeronautics and Space  
Administration under the Terrestrial Planet Finder Foundation Science  
Program.  The Lyot Project grateful acknowledges the support of the  
US Air Force and NSF in creating the special Advanced Technologies  
and Instrumentation opportunity that provides access to the AEOS  
telescope.  Eighty percent of the funds for that program are provided  
by the US Air Force.  This work is based on observations made at the  
Maui Space Surveillance System, operated by Detachment 15 of the U.S.  
Air Force Research Laboratory Directed Energy Directorate.
The Lyot Project is also grateful to the Cordelia Corporation, Hilary  
and Ethel Lipsitz, the Vincent Astor Fund, Judy Vale and an anonymous  
donor, who initiated the project. L.C.R is funded by AFRL/DE contract number F2901-00-D-0204. R.S and M.P. are supported by the 
NASA Michelson Fellowship Postdoctoral/Graduate Fellowship under contract 
to the Jet Propulsion Laboratory (JPL) funded by NASA. The JPL is managed for 
NASA by the California Institute of Technology. This work has been partially supported by
the National Science Foundation Science and Technology Center for Adaptive Optics,
managed by the University of California at Santa Cruz under cooperative
agreement AST-9876783.

\bibliography{/Users/shinkley/Desktop/papers/MasterBiblio_Sasha} 
\bibliographystyle{/Users/shinkley/Library/texmf/tex/latex/apj}

\end{document}